\documentclass[a4paper,twocolumn,superscriptaddress,10pt]{revtex4-1}

\usepackage{xcolor}
\usepackage{hyperref}
\hypersetup{
	colorlinks,
	linkcolor={red!90!black},
	citecolor={black!10!blue},
	urlcolor={blue!80!black}
}

\usepackage{tikz}
\usepackage{amssymb}
\usepackage{amsmath}
\usepackage{graphicx}
\usepackage{multirow}
\usepackage[utf8]{inputenc}
\usepackage{color}

\renewcommand{\r}{\textcolor{red}}
\newcommand{\p}{\partial}
\newcommand{\f}{\frac}
\newcommand{\md}{\mathrm{d}}

\newcommand{\br}{\mathbf{r}}
\newcommand{\fu}{\mathcal{F}^u}
\renewcommand{\u}{\mathbf{u}^{\text{ext}}}
\newcommand{\dr}{\delta\rho(\br,t)}
\newcommand{\bk}{\mathbf{k}}
\newcommand{\bq}{\mathbf{q}}
\newcommand{\drk}{\delta\rho}
\renewcommand{\theequation}{\arabic{equation}}
\renewcommand{\thefigure}{\arabic{figure}}
\renewcommand{\thesection}{\Roman{section}}
\newcommand{\Teff}{T_\text{eff}}
\renewcommand{\d}{\mathrm{d}}
\newcommand{\cp}{\chi_C^\text{peak}}
\newcommand{\TT}{\tilde{T}_\text{eff}}

\begin{document}

	\title{Active Inhomogeneous Mode-Coupling Theory (aIMCT) for Dense Systems of Self-Propelled Particles}
	\author{Soumitra Kolya}
	\email{soumitrakolya@tifrh.res.in}
	\affiliation{Tata Institute of Fundamental Research, Gopanpally Village, Hyderabad - 500046, India}
	
	\author{Puneet Pareek}
	\affiliation{Tata Institute of Fundamental Research, Gopanpally Village, Hyderabad - 500046, India}
	
    \author{Saroj Kumar Nandi}
    \email{saroj@tifrh.res.in}
    \affiliation{Tata Institute of Fundamental Research, Gopanpally Village, Hyderabad - 500046, India}
	   
\begin{abstract} 
Glassy dynamics in a dense system of active particles with self-propulsion force $f_0$ and persistence time $\tau_p$ are crucial for many biological processes. Recent studies have shown that, unlike relaxation dynamics, dynamic heterogeneity (DH) in active glasses exhibits nontrivial behavior. However, the mechanism by which activity affects DH remains unknown. We have developed an active inhomogeneous mode-coupling theory (aIMCT) for DH in active glasses. We show that the nontrivial behavior of DH comes from a novel nonequilibrium effect of activity that leads to distinct behaviors of DH and relaxation dynamics in active glasses. When activity is small, DH exhibits equilibrium-like behavior with a power-law divergence of the peak height of the four-point correlation function, $\cp$, and the aIMCT value of the exponent, $\mu\simeq 1.0$, is consistent with the existing and our new simulations of active glasses. However, $\cp$ deviates from the scaling relations at higher $f_0$ values because of the novel effect on DH, although the deviation with varying $\tau_p$ is relatively weak.

\end{abstract}
\maketitle	


{\bf Introduction:} Active glass, a novel class of  non-equilibrium dense systems {\cite{activematterreview,berthier2017,janssen2019Review}}, comprises particles with a self-propulsion force, $f_0$, and a persistence time, $\tau_p$ {\cite{ramaswamy2010, marchetti2013,vicsek2012,bechinger2016,caprini2020,cates2015b}}. Experiments \cite{fabry2001,zhou2009,garcia2015,angelini2011,sadati2013} have shown that they exhibit several characteristics of particulate glasses, such as two-step relaxation \cite{phillips1996}, non-Gaussian distribution of particle displacement \cite{chaudhuri2007}, dynamic heterogeneity \cite{franz2000}, aging \cite{cugliandolo1993,nandi2012}, etc. Examples include swimming bacteria {\cite{patteson2015,lama2024}}, ant colonies {\cite{gravish2015}}, motile cells in tissues \cite{angelini2011,garcia2015}, synthetic active matter \cite{howse2007,palacci2010,dauchot2005}, etc. Active glassy behaviors are crucial for several biological processes, such as wound healing \cite{malinverno2017, poujade2007, brugues2014}, cancer progression \cite{2009friedlb}, embryogenesis \cite{tambe2011,malmiKakkada2018}, etc. The importance of these processes demands a deeper understanding of their glassy dynamics.
Theoretical understanding of active glasses is challenging due to their non-equilibrium nature. Simulations \cite{berthier2019,flenner2016,szamel2015b} have provided crucial insights into the glassiness in these systems. Moreover, several works {\cite{berthier2019,flenner2016,szamel2014a,nandi2017b,nandi2018,cugliandolo2011,cugliandolo2019}} have shown that the departure from equilibrium is not much when activity is small {\cite{fodor2016}}, and an effective equilibrium-like description is still possible. Two well-known theories of equilibrium glasses, random first-order transition theory (RFOT) {\cite{kirkpatrick1987}} and mode-coupling theory (MCT) {\cite{gotzebook, reichman2005, das2004}}, have been extended for active systems of self-propelled particles \cite{szamel2015b,szamel2016,feng2017,liluashvili2017,nandi2017b,debets2023}. Specifically, MCT has been extended for the non-equilibrium steady-state of active glasses {\cite{berthier2013,nandi2017b,szamel2015b,paul2023}}, where the relaxation dynamics remains equilibrium-like at an effective temperature, $T_\text{eff}$ \cite{nandi2017b, nandi2018,cugliandolo2011,shen2005,szamel2014a}.

However, recent studies have shown that whereas the relaxation dynamics remains equilibrium-like, activity has nontrivial effects on the dynamic heterogeneity (DH) \cite{berthier2017,paoluzzi2022,paul2023}, and hence the material properties \cite{dhbook,ediger2000}. However, the precise mechanism of how activity affects DH is unclear. In addition, the impact of varying $\tau_p$ is also unknown. In this work, we follow the formalism of Ref. \cite{IMCT} and present the active inhomogeneous MCT (aIMCT) for the DH in active glasses. The theory shows two distinct source terms control the relaxation dynamics and DH. Hence, they may become uncorrelated. We show that the effective equilibrium-like behavior persists at low activity, although it can depart at higher activity. We have also performed molecular dynamics simulations of a well-known glassy system of active Ornstein-Uhlenbeck particles (AOUP) and compared our aIMCT predictions with simulation results.

{\bf aIMCT for active glasses:} We briefly present the aIMCT for the DH in active glasses (see supplementary materials (SM) Sec. S1 for details). Since active systems are out of equilibrium, we must write the equations of motion for both the correlation and response functions, $C_k(t)$ and $R_k(t)$, respectively, where $k$ is the wavevector. For numerical advantage, we write the equations in terms of the integrated response function, $F_k(t)=-\int_0^t R_k(t')\d t'$ \cite{nandi2017b}. Following Ref. \cite{IMCT}, we describe the DH within aIMCT in terms of the three-point-correlation function, $\chi_k^C(t)$, that has similar critical properties as the four-point correlation function, $\chi_k^4(t)$ \cite{IMCT,berthier2007a,berthier2007b}. We introduce a weak constant external field to induce a density perturbation, $\delta m_0$, and obtain the equations of motion for $\tilde{C}_k(t)$ and $\tilde{F}_k(t)$ in the presence of the field (see SM). We then define the susceptibilities as $\chi^C_k(t) =\frac{\partial {\tilde{C}}_k(t)}{\partial \delta m_0}\bigg|_{\delta m_0\to 0}$ and $\chi^F_k(t)= \frac{\partial {\tilde{F}}_k(t)}{\partial \delta m_0}\bigg|_{\delta m_0\to0}$, and obtain the equations of motion as
\begin{subequations}
	\label{chieq}
	\begin{align}
		\f{\p\chi_k^C(t)}{\p t} =&\nu_k(t)+\left(\frac{k_B T c_k}{D_L}+\zeta_k\right)C_k(t)-(T-p_k)\chi_k^C(t) \nonumber\\
		-\int_0^t &m_k(t-s)\f{\p \chi_k^C(t)}{\p s}\md s-\int_0^t \Sigma_k(t-s)\f{\p C_k(s)}{\p s}\md s \nonumber\\
		\f{\p\chi_k^F(t)}{\p t} =& \frac{k_B T c_k}{D_L} F_k(t)+\zeta_k F_k(t)-(T-p_k)\chi_k^F(t) \nonumber\\
		-\int_0^t &m_k(t-s)\f{\p \chi_k^F(t)}{\p s}\md s-\int_0^t \Sigma_k(t-s)\f{\p F_k(s)}{\p s}\md s,\nonumber
	\end{align}
\end{subequations}
where $T$ is the temperature, $\nu_k(t)=-\kappa_2^2 \int_t^\infty \Delta_k(s) \f{\p\chi_k^F(s-t)}{\p s}\md s$, $ \zeta_k=\kappa_2^2\int_0^\infty \Delta_k(s) \f{\p\chi_k^F(s)}{\p s}\md s$, and $\Delta_k(t)$ is the variance of the active noise. In this work, we consider the AOUP system, thus, $\Delta_k(t)=(f_0^2/\tau_p)\exp{(-t/\tau_p)}$ \cite{flenner2016,activematterreview}.
The memory functions are
\begin{align}
m_k(t) = \frac{\kappa_1^2}{2} &\int_{\bf q}\mathcal{V}_{k,q}^2 \f{C_q(t)C_{k-q}(t)}{T_{\text{eff}}(t)},\\
	 \Sigma_k(t)=\kappa_1^2 \int_{\bf q} &\mathcal{V}_{k,q}^2 \f{C_q(t)\chi_{k-q}^C(t)}{T_{\text{eff}}(t)} \nonumber\\
	 -\frac{\kappa_1^2}{2} &\int_{\bf q} \mathcal{V}_{k,q}^2\f{C_q(t)C_{k-q}(t)}{T_{\text{eff}}(t)^2}\frac{\p T_\text{eff}}{\p \delta m_0}\bigg|_{\delta m_0\to 0}, \label{sigmaterm}
\end{align}
with $\kappa_1=k_BT/D_Lk^2$, $\kappa_2=1/D_L$, $D_L=(\zeta+4\eta/3)/\rho_0$, and $\int_{\mathbf{q}} \equiv \int d^dq/(2\pi)^{d/2}$. The vertex function,
$\mathcal{V}_{k,q}=\bk\cdot[\bq c_q+(\bk-\bq)c_{k-q}]$, with $c_k$ being the direct correlation function. $\Teff$ comes from
\begin{equation}\label{modfdr}
	\f{\p C_k(t)}{\p t}=T_{\text{eff}}(t)\f{\p {F_k}(t)}{\p t},
\end{equation}
and 
We must solve these equations along with those for the two-point correlation functions, 
\begin{subequations}
	\begin{align}
		\f{\p C_k(t)}{\p t} &= \Pi_k(t)-(T-p_k)C_k(t)-\int_0^tm_k(t-s)\f{\p C_k(s)}{\p s}\md s , \nonumber\\
		\f{\p{F_k}(t)}{\p t} &= -1 -(T-p_k){F_k}(t)-\int_0^t m_k(t-s)\f{\p F_k(s)}{\p s}\md s,\nonumber
	\end{align}
\end{subequations}
with $\Pi_k(t) =- \kappa_2^2\int_t^\infty \Delta_k(s) \f{\p {F_k}(s-t)}{\p s}\md s$ and $p_k = \kappa_2^2 \int_0^\infty \Delta_k(s) \f{\p {F_k}(s)}{\p s}\md s$.
The second term in the source $\Sigma_k(t)$, Eq. (\ref{sigmaterm}), is proportional to $\partial \Teff/\partial \delta m_0|_{\delta m_0\to0}$ and different from equilibrium systems; it governs the nontrivial behaviors of DH in active glasses.

\begin{figure}
	\includegraphics[width=8.0cm]{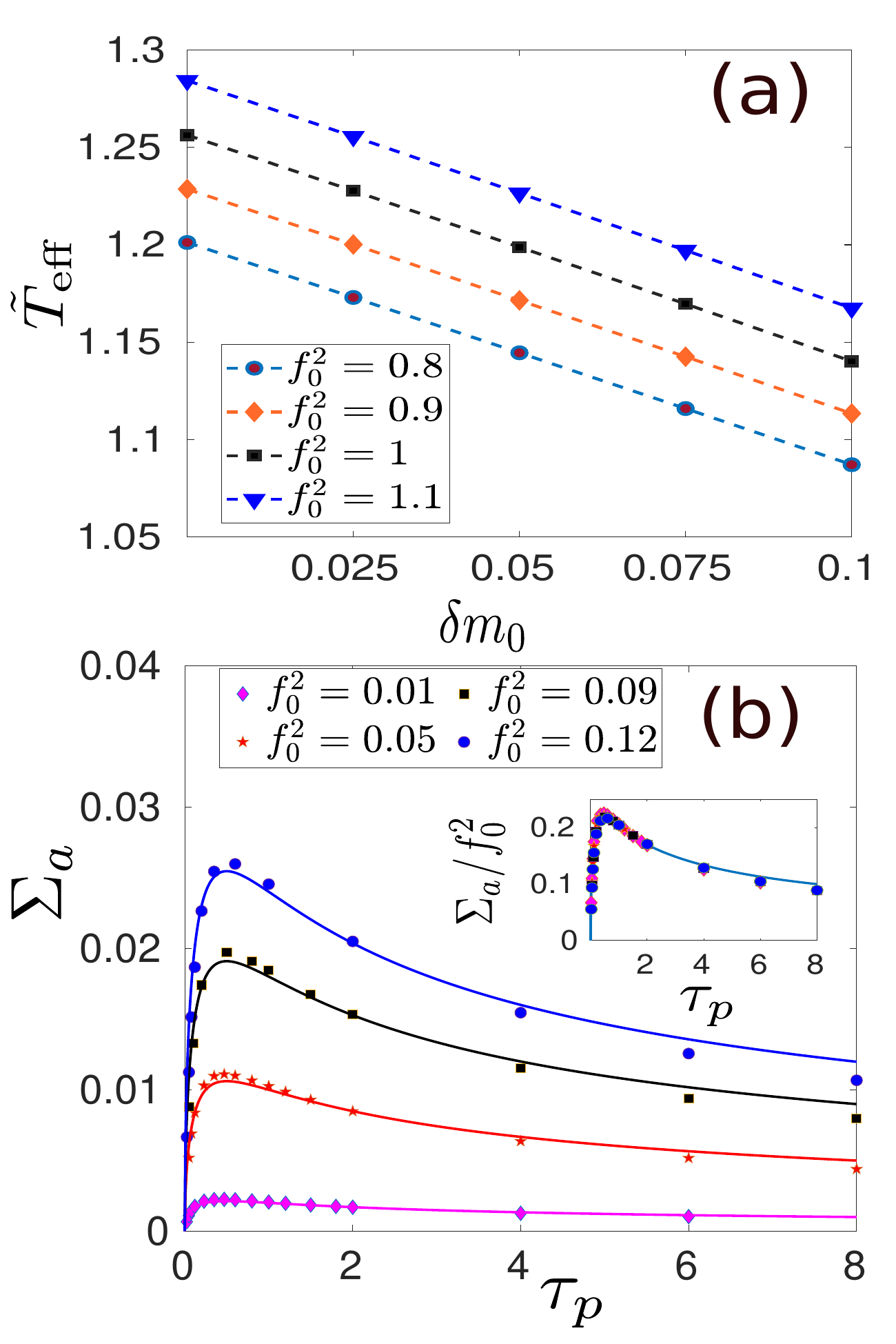}
	\caption{Behavior of $\Sigma_a$ from the numerical solution of MCT. (a) $\TT$ decays linearly at small $\delta m_0$. We have used $\lambda=1.99$. (b) Behavior of $\Sigma_a$ with $\tau_p$ for various $f_0$. Lines are fits with $k_1f_0^2\tau_p^a/(1+k_2\tau_p)$; we find that data for various $f_0$ fit well with the same values of the parameters: $k_1=0.6$, $k_2=2.0$, and $a=1/2$. {\bf Inset:} $\Sigma_a/f_0^2$ as a function of $\tau_p$ shows data collapse to a master curve. Line is the plot of $k_1\tau_p^{1/2}/(1+k_2\tau_p)$.}
	\label{activity_field_coupling}
\end{figure}

We first obtain $\Sigma_a=|\partial \Teff/\partial \delta m_0|_{\delta m_0\to0}|$ by solving the MCT equations in the presence of the external field. However, a direct solution of the full wavevector-dependent MCT equations is challenging due to enormous computational times. Therefore, we take a schematic form, simplifying the wave vector dependence by writing the equations for a single wave vector. Such simplified forms for numerical tractability have provided crucial insights under various conditions \cite{brader2009,nandi2012,gotzebook,das2004}. We present the schematic forms of the equations in the SM (Sec. S1). In the schematic version, we write $C_k(t)\equiv C(t)$, $\chi_k^C(t)\equiv \chi_C(t)$, and the memory kernels become $m(t)=2\lambda C^2(t)/\Teff(t)$ and $\Sigma(t)=4\lambda C(t)\chi_C(t)/\Teff(t)+2\lambda [C(t)^2/\Teff(t)^2] \Sigma_a$, where $\lambda$ is the control parameter.

For the calculation of $\Sigma_a$, we first calculate $\TT=[\partial \tilde{C}(t)/\partial t]/[\partial \tilde{F}/\partial t]$. In the absence of the external field, we have $\Teff=\TT|_{\delta m_0\to0}$. $\TT$ has three parts: (1) $\Teff$, (2) the contribution due to the combined effects of activity and field, and (3) a trivial part proportional to the field. The second part represents the modification due to the external field of the potential energy of a single trapped active particle in a confining potential created by the surrounding particles \cite{nandi2017b,nandi2018}. We expect this modification to be linear with $\delta m_0$ when the external field is small, such that $\delta m_0$ is also small. This modification leads to a non-zero $\Sigma_a$ for active particles. The last part in $\TT$ is just $\delta m_0$ within the schematic form. Therefore, we obtain $\Sigma_a=|(\partial \TT/\partial\delta m_0)|_{\delta m_0\to0}|-1$. Figure \ref{activity_field_coupling}(a) shows $\TT$ decreases linearly with $\delta m_0$ at small $\delta m_0$. Figure \ref{activity_field_coupling}(b) shows $\Sigma_a$ as a function of $\tau_p$ for various $f_0$ (symbols). We fit the data with a form $k_1f_0^2\tau_p^a/(1+k_2\tau_p)$ with $k_1$, $k_2$, and $a$ constants. All the data fit reasonably well with $k_1=0.6$, $k_2=2.0$, and $a=1/2$; the lines in Fig. \ref{activity_field_coupling}(b) show the plots with varying $f_0$ alone. We also find that $\Sigma_a/f_0^2$ follows data collapse to a master curve (inset of Fig. \ref{activity_field_coupling}b). Thus, we obtain
\begin{equation}\label{tempsuscep}
	\Sigma_a =\f{k_1 f_0^2\tau_p^{1/2}}{(1+k_2\tau_p)}.
\end{equation}
These specific values of $k_1=0.6$ and $k_2=2.0$ hold only within the schematic forms; we expect them to vary for different systems.

\begin{figure}
	\includegraphics[width=9cm]{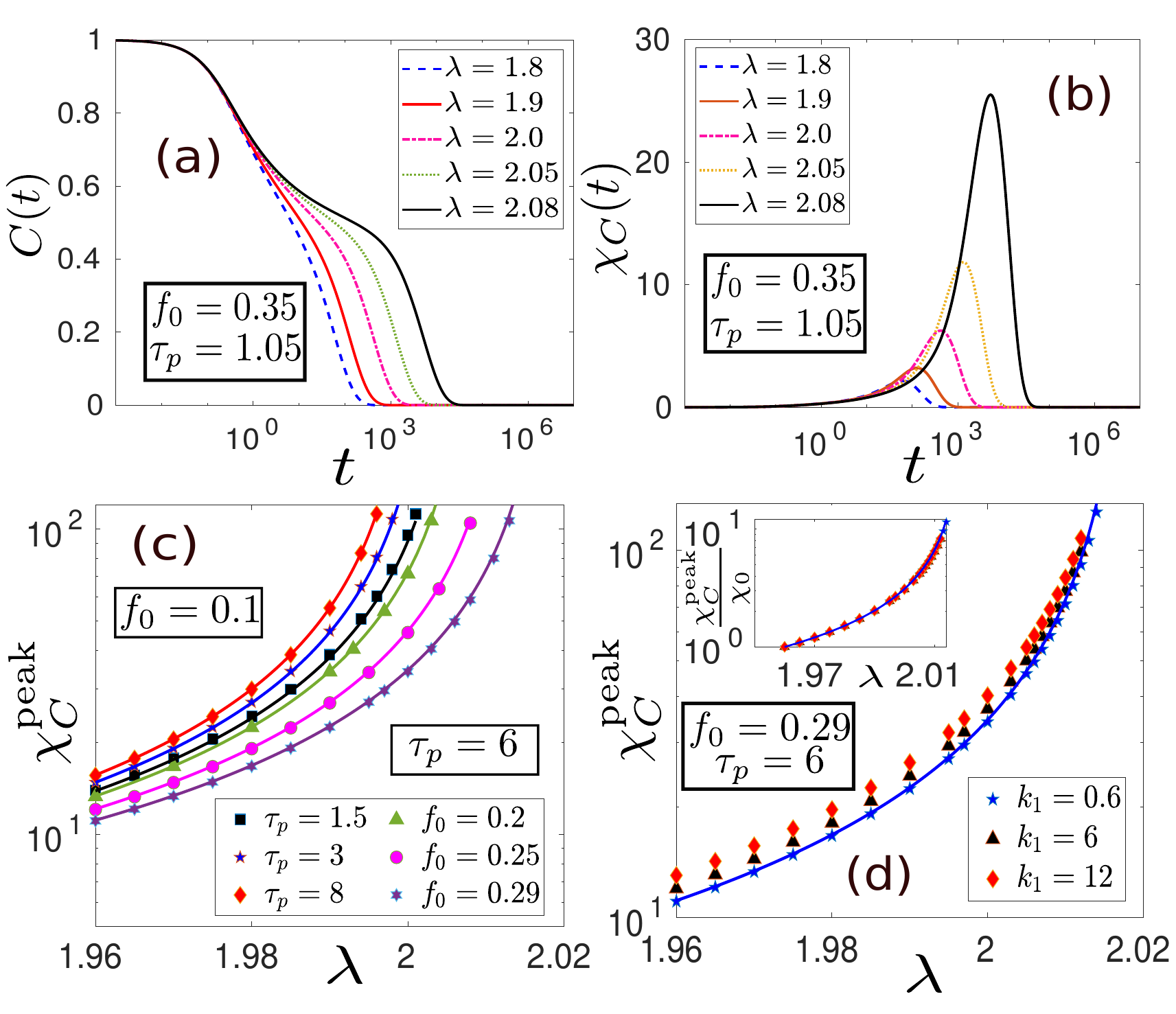}
	\caption{aIMCT results with varying $\lambda$. (a) Decay of $C(t)$ becomes slower with increasing $\lambda$. (b) The behavior of $\chi_C(t)$ for the same parameters as in (a). $\cp$ increases as the decay of $C(t)$ becomes slower. (c) $\cp$ as a function of $\lambda$ for various $f_0$ with $\tau_p=6$ and different $\tau_p$ with $f_0=0.1$. (d) $\cp$ vs. $\lambda$ for $\tau_p=6.0$ and $f_0=0.29$ with varying $k_1$. The aIMCT solutions at small $k_1$ agree with the scaling relation, Eq.(\ref{chiscaling}). {\bf Inset:} The deviation for different $k_1$ can be included in $\chi_0$. } 
	\label{mct_result_2}
\end{figure}

{\bf aIMCT results for the DH in active glasses:}
We first show the behaviors of $C(t)$ and $\chi_C(t)$ at constant activity but varying $\lambda$ (Figs. \ref{mct_result_2}a and b). Note that $\lambda=\lambda_\text{MCT}=2.0$ is the MCT transition point for the equilibrium system. Activity shifts $\lambda_\text{MCT}$ to a higher value, $\lambda_c$ \cite{berthier2013,flenner2016,mandal2016b,nandi2017b,nandi2018,activematterreview}. As $\lambda$ increases, the decay of $C(t)$ becomes slower (Fig. \ref{mct_result_2}a), and the relaxation time $\tau$, where $C(t=\tau)=0.3$, grows.
Consistent with past works \cite{flenner2016,mandal2016b,nandi2017b,nandi2018a}, the behavior of $\tau$ remains equilibrium-like at $\Teff$ (SM, Fig. S1). We expect $\chi_C(t)$ to deviate from the equilibrium-like behavior due to the $\Sigma_a$ term in Eq. (\ref{sigmaterm}). However, at constant activity, the effect of $\Sigma_a$ can be accounted for as a constant. In addition, $\Sigma_a$ is negligible at small and large $\tau_p$ (Eq. \ref{tempsuscep}). Considering the equilibrium-like behavior of $\tau$ at $\Teff$, we obtain \cite{nandi2017b},
\begin{equation} \label{tauscaling}
	\tau =\tau_0 (\lambda_c-\lambda)^{-\gamma}=\tau_0\left(\lambda_\text{MCT}-\lambda+\frac{H f_0^2}{1 + G \tau_{p}}\right)^{-\gamma}
\end{equation}
where $\tau_0$, $H$, $G$, and $\gamma$ are constants. Similarly, the scaling behavior for $\cp$ becomes
\begin{equation} \label{chiscaling}
	\cp =\chi_0 \left(\lambda_\text{MCT} -\lambda + \frac{H f_0^2}{1 + G \tau_p} \right)^{-\mu},
\end{equation}
where $\chi_0$ and $\mu$ are constants. Thus, we have $\lambda_c=\lambda_\text{MCT}+Hf_0^2/(1+G\tau_p)$. We obtain all the parameters in Eq. (\ref{tauscaling}) and (\ref{chiscaling}) by fitting the equations with a set of data of $\tau$ and $\cp$ respectively and obtain $\tau_0=4.54$, $H=2.52$, $G=1.71$, $\gamma=1.74$, $\chi_0=0.67$, and $\mu=1.0$ We can then compare the numerical solutions of aIMCT with these scaling forms, Eqs. (\ref{tauscaling}) and (\ref{chiscaling}).

Figure \ref{mct_result_2}(b) shows the characteristic non-monotonic behavior of $\chi_C(t)$: $\chi_C(t)$ first increases with $t$, reaches a peak value $\cp$ at a time $t_\text{peak}$, and then decreases at higher $t$. Figure \ref{mct_result_2}(c) shows $\cp$ becomes higher as $\lambda$ increases (symbols) and agrees well with Eq. (\ref{chiscaling}) (lines) for $k_1=0.6$. If we take $k_1$ larger, the simulation results deviate from the power-law form with constant $\chi_0$ (Fig. \ref{mct_result_2}d). However, this deviation can be included as varying $\chi_0$ such that $\cp/\chi_0$ shows data collapse to a master curve (Fig. \ref{mct_result_2}d, inset).

\begin{figure}
	\includegraphics[width=8.6cm]{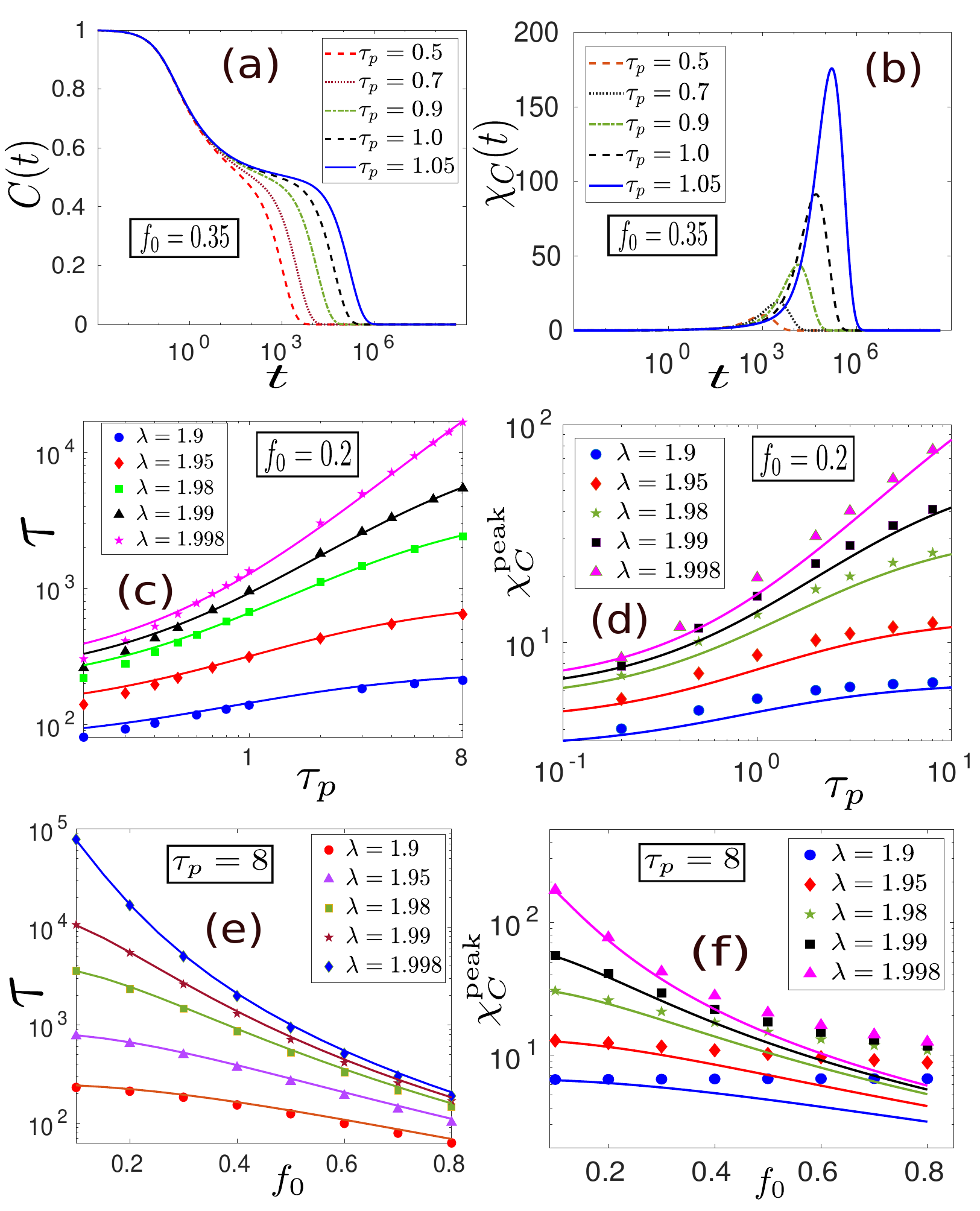}
	\caption{aIMCT results for $\tau$ and $\cp$ with $k_1=12$. (a) $C(t)$ decays slower as $\tau_p$ increases. We have used $f_0=0.35$ and $\lambda = 2.1$. (b) Corresponding behavior of $\chi_C(t)$. (c) $\tau$ grows as $\tau_p$ increases and saturates to a $\lambda$-dependent value when $\lambda<\lambda_\text{MCT}$. (d) $\cp$ also grows as $\tau_p$ increases and saturates when $\lambda<\lambda_\text{MCT}$. (e) $\tau$ decays as $f_0$ increases at constant $\tau_p$. (f) $\chi_C^\text{peak}$ decays as $f_0$ increases. In (c-f), symbols are aIMCT solutions, and the lines are the corresponding plots of either Eqs. (\ref{tauscaling}) or (\ref{chiscaling}).}
	\label{mct_result_1}
\end{figure}

Next, we discuss the behavior with varying activity. We expect a deviation of $\cp$ from the effective-equilibrium behavior. $k_1$ sets a scale for $f_0$, and we expect it to vary for different systems. The deviation is not significant at $k_1 = 0.6$ as $\Sigma_a$ is small (Fig. S2 in the SM). Therefore, to highlight the deviation, we take $k_1=12$ for these solutions. Figure \ref{mct_result_1}(a) shows the decay of $C(t)$ at $\lambda=2.1$ and $\tau_p$. $C(t)$ decays slower as $\tau_p$ increases as expected for the AOUPs \cite{flenner2016}. Figure \ref{mct_result_1}(b) shows the corresponding behavior of $\chi_C(t)$. $\cp$ increases as the decay of $C(t)$ becomes slower. $\tau$ and $\cp$ increase with $\tau_p$ and saturate as $\tau_p\to\infty$ when $\lambda<\lambda_c$ (Figs. \ref{mct_result_1}c and d). The lines show the plots of Eqs. (\ref{tauscaling}) and (\ref{chiscaling}) with the same parameters as before.

We have also studied the behavior of $\tau$ and $\cp$ as functions of $f_0$ [Figs. \ref{mct_result_1}(e) and (f)]: the symbols are the numerical solutions of the aIMCT and lines are the scaling relations, Eqs. (\ref{tauscaling}) and (\ref{chiscaling}). Whereas $\tau$ and $\cp$ agree remarkably well at lower activity with the scaling relations, $\cp$ deviates significantly at higher $f_0$. We can understand this deviation of $\cp$ using Eq. (\ref{tempsuscep}); this term is negligible at small activity and becomes significant only at higher $f_0$. However, the agreement with varying $\tau_p$ is much better as $\Sigma_a\to0$ for both small and large $\tau_p$. The deviation from Eq. (\ref{chiscaling}) is significant only at intermediate $\tau_p$ (Fig. \ref{mct_result_1}d). These results demonstrate the effects of two distinct terms controlling the behaviors of $\tau$ and $\chi_C(t)$. When activity is small, the deviation from equilibrium-like behavior is not significant. However, at higher activity, the behaviors of relaxation dynamics and DH decouple as two distinct source terms control their behaviors.

\begin{figure}
	\includegraphics[width=8.6cm]{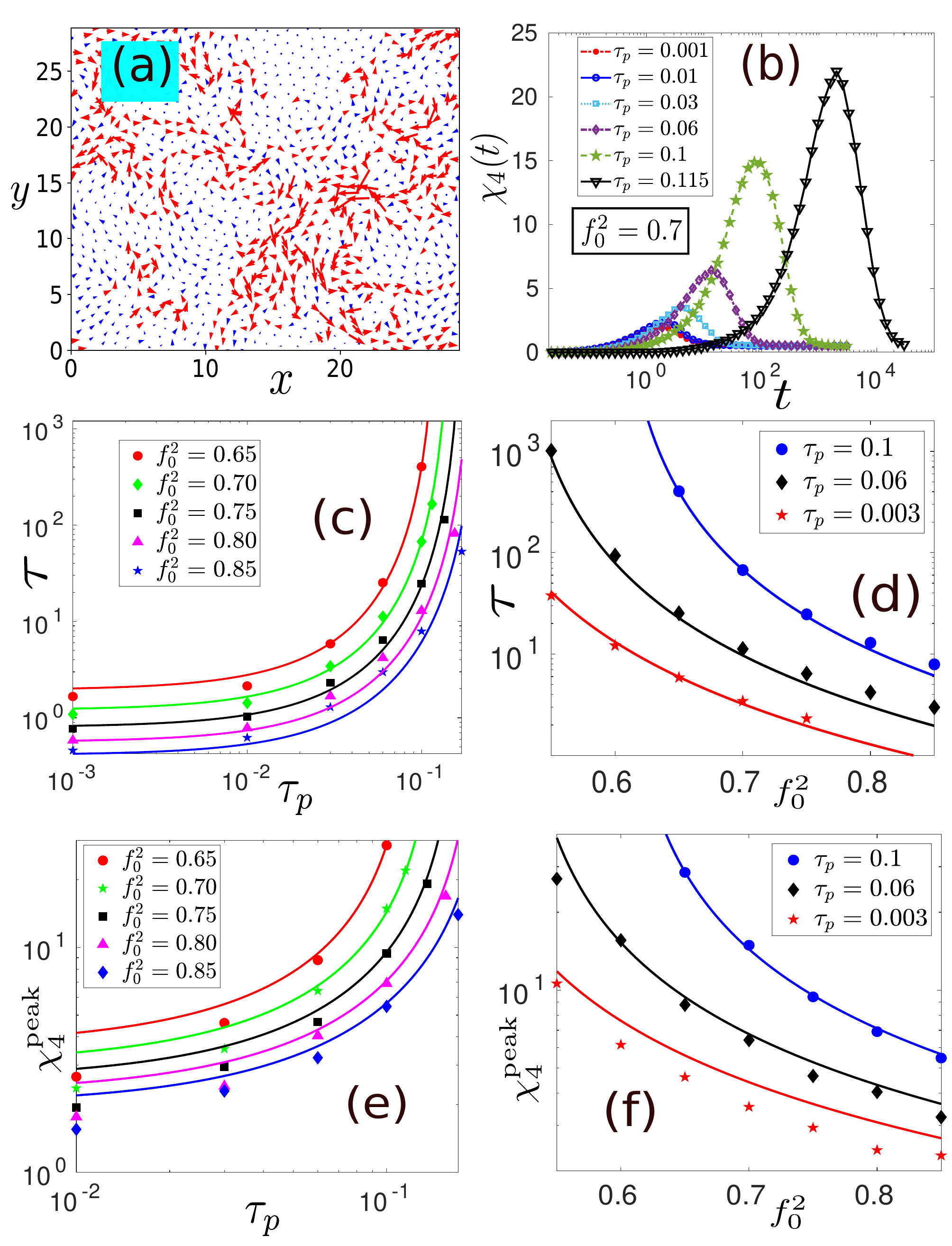}
	\caption{Comparison with simulation. (a) Illustration of dynamic heterogeneity in a two-dimensional ($2d$) simulation for better visibility. Arrow directions show the velocity and lengths are proportional to the speed. Red denotes particles with a velocity higher than the average, and blue denotes those with lower velocities. (b) $\chi_4(t)$ peak grows as $\tau_p$ increases and shifts to higher $t$. (c) $\tau$ grows as $\tau_p$ increases. (d) $\tau$ decreases as $f_0$ increases. The symbols are simulation data, and lines are the plots of the scaling relation with $\tau_0 = 0.057$, $H = 1.05$, $G = 5.14$, $\gamma = 2.69$ and $T_\text{MCT} = 0.41$. (e) $\chi_4^\text{peak}$ grows as $\tau_p$ increases. (f) $\chi_4^\text{peak}$ decays as $f_0$ increases. The symbols are simulation data, and lines are the scaling relations with $\chi_0 = 0.92$ and $\mu = 1.04$. } 
	\label{comparision_with_simulation}
\end{figure}

{\bf Comparison with simulations:} Finally, we compare the aIMCT predictions with our simulations of AOUPs \cite{flenner2016}. To highlight the effects of activity, we have simulated an athermal system. Although our simulations are in $3d$, we show the real space image of DH for a $2d$ system for ease of presentation (Fig.  \ref{comparision_with_simulation}a); similar to other glassy systems \cite{activematterreview}, the fast and slow-moving particles tend to cluster together. In our simulations, we quantify DH via the four-point correlation function (defined in the SM) that has identical critical properties as that of $\chi_C(t)$ \cite{berthier2007a,IMCT}. The peak height, $\chi_4^\text{peak}$, of $\chi_4(t)$ grows and shifts to longer times as $\tau_p$ increases (Fig. \ref{comparision_with_simulation}b); this is very similar to the behavior of $\chi_C(t)$ (Fig. \ref{mct_result_1}b).

The scaling relations, Eqs. (\ref{tauscaling}) and (\ref{chiscaling}), for the athermal system becomes $\tau=\tau_0 \epsilon^{-\gamma}$ and $\chi_4^\text{peak}=\chi_0\epsilon^{-\mu}$, where $\epsilon={H f_0^2}/(1 + G \tau_p)-T_\text{MCT}$. Figures \ref{comparision_with_simulation}(c) and (d) show the simulation data (symbols) of $\tau$ with varying $\tau_p$ and $f_0$, respectively. Figures \ref{comparision_with_simulation}(e) and (f) show the corresponding simulation data (symbols) for $\chi_4^\text{peak}$. The lines are comparisons with the MCT scaling relations with the same set of parameters with $\gamma=2.7$ and $\mu=1.04$. Note that the value of $\mu$ is very close to that in aIMCT. Similar to the aIMCT results, the agreement for the $\chi_4^\text{peak}$ becomes worse at larger $f_0$, although the data of $\tau$ agree well with the scaling relations. In addition, the agreement with varying $\tau_p$ is relatively better. aIMCT explains these behaviors of DH through the nonequilibrium effects of activity via the specific form of $\Sigma_a$, Eq. (\ref{tempsuscep}).

{\bf Conclusion:}
Recent works have demonstrated that the DH in active glasses has qualitatively different behavior compared to the relaxation dynamics \cite{berthier2017,paoluzzi2022,paul2023}. We have analytically derived the active inhomogeneous mode-coupling theory (aIMCT) for the DH in active glasses. aIMCT traces the different behaviors of DH and relaxation dynamics in active glasses to distinct source terms that control their behaviors. We have calculated the activity-specific nonequilibrium contribution, $\Sigma_a$, by solving the MCT under a perturbing field. For AOUP systems, $\Sigma_a\propto f_0^2\tau_p^{1/2}/(1+k_2 \tau_p)$. Thus, the behavior of DH at large $f_0$ will deviate from the equilibrium-like behavior. However, the deviation for varying $\tau_p$ is much less as $\Sigma_a$ has a maximum with varying $\tau_p$. Using the equivalence of active Brownian particles (ABP) \cite{debets2021,debets2022}, we expect $\Sigma_a \propto f_0^2\tau_p^{3/2}/(1+k_2 \tau_p)$ for ABPs. In this case, the deviation from equilibrium-like behavior will also be higher with varying $\tau_p$, although the behavior for varying $f_0$ will be similar within both models.

The non-zero value of $\Sigma_a$ is a strictly nonequilibrium effect and comes from the fact that $\Teff$ represents a potential energy rather than a temperature-like variable. Thus, although $\Teff$ continues to play a crucial role in understanding the behavior in systems out of equilibrium \cite{nandi2018,cugliandolo2011,cugliandolo2019}, we should be cautious in its interpretation.
Nevertheless, the effect of activity via $\Sigma_a$ will be crucial for systems under external field or pinning when the glassy system is driven towards the higher-order critical point \cite{nandi2014}. Interestingly, the DH exponent in simulations is very close to that of aIMCT in a regime where the effective equilibrium behavior persists. As the behavior of DH bears a clear signature of nonequilibrium effects of activity, our analytical theory for DH should be instrumental for deeper insights into how active fluctuations affect glassy dynamics.

{\bf Acknowledgments:} We acknowledge the support of the Department of Atomic Energy, Government of India, under Project Identification No. RTI 4007. SKN thanks SERB for grant via SRG/2021/002014.


%



\pagebreak
\newpage
\widetext
\begin{center}
	\textbf{\large{S\lowercase{upplementary} M\lowercase{aterial}: A\lowercase{ctive} I\lowercase{nhomogeneous} M\lowercase{ode}-C\lowercase{oupling} T\lowercase{heory} (\lowercase{a}IMCT) \lowercase{for} D\lowercase{ense} S\lowercase{ystems} of S\lowercase{elf}-P\lowercase{ropelled} P\lowercase{articles}}}
\end{center}

\setcounter{table}{0}
\setcounter{figure}{0}
\setcounter{section}{0}
\setcounter{equation}{0}
\renewcommand{\thefigure}{S\arabic{figure}}
\renewcommand{\thetable}{S\arabic{table}}
\renewcommand{\theequation}{S\arabic{equation}}
\renewcommand{\thepage}{S\arabic{page}}
\renewcommand{\thesubsection}{S\arabic{subsection}}
\renewcommand{\bibnumfmt}[1]{[S#1]}
\renewcommand{\citenumfont}[1]{S#1}
\renewcommand{\d}{\mathrm{d}}
\renewcommand{\r}{\textcolor{red}}
\renewcommand{\u}{\mathbf{u}^{\text{ext}}}
\renewcommand{\theequation}{\arabic{equation}}
\renewcommand{\thefigure}{\arabic{figure}}
\renewcommand{\thesection}{\Roman{section}}
\newcommand{\D}{\mathcal{D}}
\renewcommand{\a}{\alpha}
\renewcommand{\b}{\beta}
\renewcommand{\d}{\mathrm{d}}
\renewcommand{\t}{\tau}
\renewcommand{\S}{\Sigma}
\renewcommand{\l}{\lambda}
\newcommand{\gdot}{\dot{\gamma}}
\renewcommand{\tt}{\tilde{\tau}}
\newcommand{\e}{\epsilon}
\newcommand{\DD}{\Delta}
\newcommand{\Tsp}{T_{eff}^{sp}}
\newcommand{\bv}{\mathbf{v}}
\newcommand{\nf}{\mathbf{f}}
\newcommand{\mF}{\mathcal{F}}

\subsection{Active Inhomogeneous Mode-Coupling Theory (aIMCT)}
\label{MCTdetails}
A direct calculation of the four-point correlation function within mode-coupling theory (MCT) is challenging. Biroli {\it et al.} introduced the inhomogeneous mode-coupling theory (IMCT) formalism \cite{IMCT-1} to analyze dynamical heterogeneity (DH) in terms of a related three-point correlation function that has similar critical properties as that of the four-point correlation function. IMCT obtains the three-point correlation function via a susceptibility.
We start with the hydrodynamic equations of motion for $\rho(\br,t)$ and the momentum density, $\rho(\br,t)\mathbf{v}(\br,t)$, for an active fluid:
\begin{align}
	\f{\p\rho(\br,t)}{\p t}&=-\nabla\cdot (\rho(\br,t)\mathbf{v}(\br,t)) \label{contrho}\\
	\f{\p (\rho(\br,t)\mathbf{v}(\br,t))}{\p t}+\nabla\cdot(\rho(\br,t)\mathbf{v}(\br,t)\mathbf{v}(\br,t)) &=\eta \nabla^2\mathbf{v}(\br,t)+\left(\zeta+\f{\eta}{3}\right)\nabla\nabla\cdot\mathbf{v}(\br,t)-\rho(\br,t)\nabla\f{\delta \fu}{\delta\rho(\br,t)}+\mathbf{f}_T(\br,t)+\mathbf{f}_A(\br,t), \label{contmo}
\end{align}
where $\eta$ and $\zeta$ are bulk and shear viscosities, respectively, and $\mathbf{f}_T(\br,t)$ is the thermal noise with zero mean and variance as
\begin{equation}
	\langle \mathbf{f}_T(\mathbf{0},0)\mathbf{f}_T(\br,t)\rangle=-2k_BT\left[ \eta \mathbf{I}\nabla^2+\left(\zeta+\f{\eta}{3}\right)\nabla\nabla \right]\delta(\br)\delta(t)
\end{equation}
with $k_BT$ being the Boltzmann constant times temperature, $\mathbf{I}$, the unit tensor in spatial dimension $d$. Activity in our system enters via the active noise, $\mathbf{f}_A(\br,t)$, with zero mean and 
\begin{equation}
	\langle \mathbf{f}_{A}(\br,t)\mathbf{f}_A(\mathbf{0},0)\rangle=2\Delta(\br,t)\mathbf{I},
\end{equation}
where $\Delta(\br,t)$ specifies the type of activity. In this work, we have focused on the active Ornsterin-Uhlenbeck particles (AOUPs). $\fu$ is a free energy functional in the presence of $\u$
\begin{align}
	\beta\fu=\int_{\br}\left[\rho(\br,t)\ln\left(\f{\rho(\br,t)}{\rho_0}\right)-\delta\rho(\br,t)\right]-\f{1}{2}\int_{\br,\br'} c(\br-\br')\delta\rho(\br,t)\delta\rho(\br',t)+\beta\int_\br \u(\br)\delta\rho(\br,t),
\end{align}
where $\beta=1/k_BT$. In the presence of of $\u(\br)$, the static density, $m(\br)$, becomes inhomogeneous and space-dependent. Therefore, we have $\rho(\br,t)=m(\br)+\dr$, where $\dr$ denotes the fluctuating part of density. We have used the notation $\int_\br\equiv \int \md\br$. We obtain $m(\br)$ by minimizing $\fu$:
\begin{equation}
	\beta\f{\delta\fu}{\dr}\bigg|_{\rho(\br,t)=m(\br)}=0.
\end{equation}

We now divide the static inhomogeneous density into two contributions, $\rho_0$, the homogeneous background density, and $\delta m(\br)$, the static density due to the presence of $\u$. Thus, $m(\br)=\rho+\delta m(\br)$, where $\delta m(\br)$ is small as $\u$ is weak. The density fluctuation, $\dr$, and the velocity field are small in the glassy regime. Therefore, we can linearize Eqs. (\ref{contrho}) and (\ref{contmo}), by neglective $\dr\mathbf{v}(\br,t)$ and higher-order terms. We disregard both the acceleration and inertia terms in Eq. (\ref{contmo}), take a divergence and replace $\nabla\cdot\mathbf{v}(\br,t)$ using Eq. (\ref{contrho}). This necessitates the evaluation of $\nabla\cdot\left(\rho(\br,t)\nabla(\delta \beta\fu/\dr)\right)$. We obtain the force density, within linear order in $\delta m(\br)$, as
\begin{align}
	\rho(\br,t)\nabla\f{\delta \fu}{\dr}&=\nabla\int_\br[\delta(\br-\br')-\rho_0c(r-r')]\delta\rho(\br',t)-\nabla\int_{\br'}\delta m(\br)c(r-r')\delta\rho(\br',t) \nonumber\\
	&-\dr\nabla\int_{\br'}c(r-r')\delta\rho(\br',t)-\f{\nabla\delta m(\br)}{\rho_0}[\delta(\br-\br')-\rho_0c(r-r')]\delta\rho(\br',t).
\end{align}
We write down the divergence of the force density in Fourier-space, at wavevector $\bk$, as
\begin{align}\label{divforce}
	\left[\nabla\cdot\left(\rho(\br,t)\nabla\f{\delta\beta\fu}{\dr}\right)\right]_\bk&=-k^2k_BT (1-\rho_0c_k)\drk_\bk(t)+k^2k_BT\int_\bq\delta m_{\bk-\bq}c_q\drk_\bq(t) \nonumber\\
	&+\f{k_BT}{\rho_0}\int_\bq\frac{\bk\cdot(\bk-\bq)}{S_q}\delta m_{\bk-\bq}\drk_\bq(t)+\f{k_BT}{2}\int_\bq \bk\cdot[\bq c_q+(\bk-\bq)c_{k-q}]\drk_\bq(t)\drk_{\bk-\bq}(t),
\end{align}
where $c_k$ and $S_k$ are the direct correlation function and static structure factor, respectively. Within MCT, the glass transition is a critical phenomenon with a diverging length scale. 
Therefore, we take a constant $\u$ such that $\delta m_\bk$ is sharply peaked at around $\bk\to0$. Then, Eq. (\ref{divforce}) for this particular choice of $\u$ becomes
\begin{align}
	\left[\nabla\cdot\left(\rho(\br,t)\nabla\f{\delta\beta\fu}{\dr}\right)\right]_\bk=-\f{k^2k_BT}{S_k}\drk_\bk(t)+k^2k_BT\delta m_{\bf 0}c_k\drk_\bk(t)+\f{k_BT}{2}\int_\bq \bk\cdot[\bq c_q+(\bk-\bq)c_{k-q}]\drk_\bq(t)\drk_{\bk-\bq}(t).
\end{align}
Using the above expression, we obtain the equation of motion for density fluctuation in Fourier space as
\begin{align}\label{lankspace}
	D_Lk^2\f{\p\drk_\bk(t)}{\p t}&+\f{k_BTk^2}{S_k}\drk_\bk(t)-k^2k_BT\delta m_{\bf 0}c_k\drk_\bk(t) \nonumber\\
	&=\f{k_BT}{2}\int_\bq \bk\cdot[\bq c_q+(\bk-\bq)c_{k-q}]\drk_\bq(t)\drk_{\bk-\bq}(t)+ik\hat{f}_T^L(\bk,t)+ik\hat{f}_A^L(\bk,t),
\end{align}

where $\mathcal{V}_{k,q}=\bk\cdot[\bq c_q+(\bk-\bq)c_{k-q}]$, $\hat{f}_T^L$ and $\hat{f}_A^L$ are the longitudinal parts of the Fourier transforms of $\nf_T$ and $\nf_A$, $D_L=(\zeta+4\eta/3)/\rho_0$. $S_k=1/(1-\rho_0c_k)$.
Through a field-theoretical method \cite{reichman2005-1,nandi2016-1},
we obtain the equations for the correlation, $\tilde{C}_k(t,t')=\langle\delta\rho_k(t)\delta\rho_{-k}(t')\rangle$, and the response, $\tilde{R}_k(t,t')=\langle \p\delta\rho_k(t)/\p \hat{f}_T^L(t')\rangle$, functions (the tilde denotes they are defined in the presence of the field) as
\begin{align}\label{kdepeq1}
	\f{\p \tilde{C}_k(t,t')}{\p t} &=-\mu_k(t)\tilde{C}_k(t,t') +\frac{k_BTc_k\delta m_0}{D_L}\tilde{C}_{k}(t,t') + \int_0^{t'}\d s\D_k(t,s)\tilde{R}(t',s) 
	+\int_0^t\d s \S_k(t,s)\tilde{C}_k(s,t')+2T\tilde{R}_k(t',t)\\
	\f{\p \tilde{R}_k(t,t')}{\p t} &=-\mu_k(t)\tilde{R}_k(t,t') +\frac{k_BTc_k\delta m_0}{D_L}\tilde{R}_{k}(t,t')
	+\int_{t'}^t \d s\S_k(t,s)\tilde{R}_k(s,t')+\delta(t-t') \\
	\mu_k(t) = T&\tilde{R}_k(0)+\int_0^t \d s[\D_k(t,s)\tilde{R}_k(t,s)+\S_k(t,s)\tilde{C}_k(t,s)] \nonumber
\end{align}

with,
\begin{align}
	\Sigma_k(t,s)&= \kappa_1^2 \int_{\bf q} \mathcal{V}_{k,q}^2 \tilde{C}_{k-q}(t,s)\tilde{R}_q(t,s),  \nonumber \\
	\D_k(t,s)&=\f{\kappa_1^2}{2} \int_{\bf q} \mathcal{V}_{k,q}^2 \tilde{C}_q(t,s)\tilde{C}_{k-q}(t,s)+\kappa_2^2\Delta_k(t-s), \nonumber
\end{align}
where $\kappa_1=k_BT/D_Lk^2$ and $\kappa_2=1/D_L$.
We now define the susceptibilities
\cite{IMCT-1} corresponding to ${C}_k$ and ${R}_k$ as: $\chi^C_k(t,t') =
{\partial {\tilde{C}}_k(t,t')}/{\partial \delta m_0}|_{\delta m_0\to 0}$ and
$\chi^R_k(t,t')= {\partial {\tilde{R}}_k(t,t_w)}/{\partial \delta m_0}|_{\delta
	m_0\to0}$. Then, using Eq. (\ref{kdepeq1}), we obtain the equations of motion as

\begin{subequations}
	\begin{align}
		\label{chiRCa}
		\frac{\partial \chi^{C}_k(t,t')}{\partial t}& +\mu_k(t)\chi^{C}_k(t,t') 
		=\int_0^{t'}\d s {\D}_k(t,s)\chi^{R}_k(t',s) 
		+\int_0^{t'}\d s  {\D}'_k(t,s){R}_k(t',s) \\ \nonumber&+\int_{0}^t \d s {\Sigma}_k(t,s)\chi^{C}_k(s,t') 
		+\int_{0}^t \d s {\Sigma}'_k(t,s){C}_k(s,t') +\mathcal{S}^C_k(t,t'),\\
		\label{chiRCb}
		\frac{\partial \chi_k^{R}(t,t')}{\partial t}& + \mu_k(t)\chi_k^{R}(t,t') 
		=\int_{t'}^t \d s {\Sigma}_k(t,s)\chi^{R}_k(s,t') 
		+\int_{t'}^t \d s {\Sigma}'_k(t,s){R}_k(s,t')
		+\mathcal{S}^R_k(t,t'), \\		
	\end{align}
\end{subequations}

where ${\Sigma}'_k(t,s)=\partial {\Sigma}_k(t,s)/\partial\delta
m_0|_{\delta m_0\to0}$,  and ${\D}'_k(t,s)={\partial
	{\D}_k(t,s)}/{\partial\delta m_0}|_{\delta m_0\to0}$. The expressions for the source terms $\mathcal{S}^R_k(t,t')$ and
$\mathcal{S}^C_k(t,t')$ are given in the form
\begin{align}
	\mathcal{S}_k^R(t,t')=\frac{k_BTc_k}{D_L}R_k(t,t')-\omega_k(t)R_k(t,t'),
	\nonumber\\
	\mathcal{S}_k^C(t,t')=\frac{k_BTc_k}{D_L}C_k(t,t')-\omega_k(t)C_k(t,t')
\end{align}
with 
\begin{align}
	\omega_k(t)=&\kappa_1^2\int_0^t \int_{\bf q} \mathcal{V}_{k,q}^2\bigg[\chi^C_{k-q}(t,s)\bigg\{\frac{1}{2}C_q(t,s)R_k(t,s)\nonumber+R_q(t,s)C_k(t,s)\bigg\}\d s \nonumber\\
	+&C_{k-q}(t,s)\bigg\{\frac{1}{2}\chi^C_q(t,s)R_k(t,s)+\frac{1}{2}C_q(t,s)\chi^R_k(t,s) \chi^R_q(t,s)C_k(t,s) \\ &+R_q(t,s)\chi_k^C(t,s)\bigg\}\bigg]\d s + \kappa_2^2\int_0^t \Delta(t-s) \chi_k^R(t,s) \d s .
\end{align}

For the advantage of numerical solutions, we define an integrated response function and the corresponding susceptibility, ${F}(t,t')$ and $\chi_F(t,t')$, as
\begin{equation}
	{F_k}(t,t')=-\int_{t'}^t{R_k}(t,s)\md s;\,\,\, {\chi_k^F}(t,t')=-\int_{t'}^t{\chi_k^R}(t,s)\md s.
\end{equation}

Without the loss of generality, we consider $t>t'$, and therefore ${R}(t',t)$ becomes zero due to the boundary condition obeying causality. We are interested in the steady state dynamics of the active system. Therefore, we take the limit $t\to\infty$ and $t'\to\infty$ such that $(t-t')$ remains constant, then $C_k(t,t')$, ${F_k}(t,t')$, $\chi_k^C(t,t')$ and $\chi_k^F(t,t')$ become functions of the time difference alone, i.e., 
\begin{equation}
	C_k(t,t')\equiv C_k(t-t'); \,\,\,  {F_k}(t,t')\equiv {F_k}(t-t'); \,\,\, 	\chi_k^C(t,t')\equiv \chi_k^C(t-t'); \,\,\,  \chi_k^{F}(t,t')\equiv \chi_k^{F}(t-t').
\end{equation}
We redefine $(t-t')\to t$ and 
obtain the equations of motions for $C_k(t)$ and $F_k(t)$  as

\begin{subequations}
	\begin{align}
		\f{\p C_k(t)}{\p t} &= \Pi_k(t)-(T-p_k)C_k(t)-\int_0^tm_k(t-s)\f{\p C_k(s)}{\p s}\md s + \frac{k_B T c_k}{D_L} \delta m_0 C_k(t),\\
		\f{\p{F_k}(t)}{\p t} &= -1 -(T-p_k){F_k}(t)-\int_0^t m_k(t-s)\f{\p F_k(s)}{\p s}\md s + \frac{k_B T c_k}{D_L} \delta m_0 F_k(t),
	\end{align}
\end{subequations}
with,
\begin{align}
	&\Pi_k(t) =- \kappa_2^2\int_t^\infty \Delta_k(s) \f{\p {F_k}(s-t)}{\p s}\md s,\\
	&p_k = \kappa_2^2 \int_0^\infty \Delta_k(s)\f{\p {F_k}(s)}{\p s}\md s, \\
	&m_k(t-s) = \frac{\kappa_1^2}{2} \int_{\bf q} \mathcal{V}_{k,q}^2 \f{C_q(t-s)C_{k-q}(t-s)}{T_{\text{eff}}(t-s)},
	\label{teff_final}
\end{align}

where $T_{\text{eff}}(t)$ is defined through a generalized fluctuation-dissipation relation (FDR) for nonequilibrium systems as
\begin{equation}
	\f{\p C_k(t)}{\p t}=T_{\text{eff}}(t)\f{\p {F_k}(t)}{\p t}. 
\end{equation}

Similarly, we get the steady state equations of $\chi_k^C(t) $ and $\chi_k^F(t)$ as follows

\begin{subequations}
	\label{chieq}
	\begin{align}
		\f{\p\chi_k^C(t)}{\p t} &=\nu_k(t)+\frac{k_B T c_k}{D_L}C_k(t)+\zeta_kC_k(t)-(T-p_k)\chi_k^C(t)-\int_0^tm_k(t-s)\f{\p \chi_k^C(t)}{\p s}\md s-\int_0^t \Sigma_k(t-s)\f{\p C_k(s)}{\p s}\md s \\
		\f{\p\chi_k^F(t)}{\p t} &=\frac{k_B T c_k}{D_L} F_k(t)+\zeta_k F_k(t)-(T-p_k)\chi_k^F(t)-\int_0^tm_k(t-s)\f{\p \chi_k^F(t)}{\p s}\md s-\int_0^t \Sigma_k(t-s)\f{\p F_k(s)}{\p s}\md s,
	\end{align}
\end{subequations}

where
\begin{align}
	\nu_k(t)&=-\kappa_2^2 \int_t^\infty \Delta_k(s) \f{\p\chi_k^F(s-t)}{\p s}\md s \\
	\zeta_k&=\kappa_2^2\int_0^\infty \Delta_k(s) \f{\p\chi_k^F(s)}{\p s}\md s\\
	\text{and} \,\,\, \Sigma_k(t)&=\kappa_1^2 \int_{\bf q} \mathcal{V}_{k,q}^2 \f{C_q(t)\chi_{k-q}^C(t)}{T_{\text{eff}}(t)}-\frac{\kappa_1^2}{2} \int_{\bf q} \mathcal{V}_{k,q}^2\f{C_q(t)C_{k-q}(t)}{T_{\text{eff}}(t)^2}\frac{\p T_\text{eff}}{\p \delta m_0}, \label{sigmaterm}
\end{align}

We take a schematic approximation, writing the theory at a particular wave vector $k_{max}$, which corresponds to the first maximum of the static structure factor, that leads to simplified equations manageable for numerical solution. Then we obtain the equations for $\chi_C(t) = \chi_{k = k_{max}}^C(t)$ , $\chi_F(t) = \chi_{k = k_{max}}^F(t)$, $C(t)\equiv C_{k=k_{max}}(t)$ and $R(t) \equiv R_{k=k_{max}}(t)$ in zero field limit as

\begin{subequations}
	\label{chieq}
	\begin{align}
		\f{\p\chi_C(t)}{\p t} &=\nu(t)+(1+\zeta)C(t)-(T-p)\chi_C(t)-\int_0^tm(t-s)\f{\p \chi_C(t)}{\p s}\md s-\int_0^t \Sigma(t-s)\f{\p C(s)}{\p s}\md s \\
		\f{\p\chi_F(t)}{\p t} &=(1+\zeta)F(t)-(T-p)\chi_F(t)-\int_0^tm(t-s)\f{\p \chi_F(t)}{\p s}\md s-\int_0^t \Sigma(t-s)\f{\p F(s)}{\p s}\md s,
	\end{align}
\end{subequations}
where
\begin{align}
	\nu(t)&=-\int_t^\infty \Delta(s) \f{\p\chi_F(s-t)}{\p s}\md s \\
	\zeta&=\int_0^\infty \Delta(s) \f{\p\chi_F(s)}{\p s}\md s\\
	\text{and} \,\,\, \Sigma(t)&=4\lambda\f{C(t)\chi_C(t)}{T_{\text{eff}}(t)}-2\lambda\f{C(t)^2}{T_{\text{eff}}(t)^2}\frac{\p T_\text{eff}}{\p \epsilon}, \label{sigmaterm}
\end{align}
along with the equations for the two-point correlation functions
\begin{subequations}
	\begin{align}
		\f{\p C(t)}{\p t} &= \Pi(t)-(T-p)C(t)-\int_0^tm(t-s)\f{\p C(s)}{\p s}\md s\\
		\f{\p{F}(t)}{\p t} &= -1 -(T-P){F}(t)-\int_0^t m(t-s)\f{\p F(s)}{\p s}\md s,
	\end{align}
\end{subequations}
with,
\begin{align}
	&\Pi(t) =-\int_t^\infty \Delta(s) \f{\p {F}(s-t)}{\p s}\md s,\\
	&p = \int_0^\infty \Delta(s)\f{\p {F}(s)}{\p s}\md s \\
	&m(t-s) =2\lambda \f{C^2(t-s)}{T_{\text{eff}}(t-s)}
	\label{teff_final}
\end{align}
where $\lambda$ is the control parameter of the passive system. An analytical solution of the IMCT equations is not possible, and one must solve them numerically. Even then, the numerical solution of the full wavevector-dependent theory is impractical as it requires a long computational time. Therefore, we have solved the schematic version of the theory.

\subsection{Simulation Details}
We simulated the Kob-Anderson binary mixture \cite{kob1995-1} of 80:20 in 3d and 65:35 in 2d. The interaction potential is given by
\begin{equation}
	V_{\alpha\beta}(r) = 4\epsilon_{\alpha\beta}[(\sigma_{\alpha\beta}/r)^{12}-(\sigma_{\alpha\beta}/r)^{6}]
\end{equation}
if $r > 2.5\sigma_{\alpha\beta}$ and $0$ otherwise. Here $\alpha,\beta$ signifies the type of the particle (A or B), and $r$ is the distance between two particles. The particle density is 1.2 and the parameters are $\epsilon_{AA} = 1.0$, $\sigma_{AA} = 1.0$, $\epsilon_{AB} = 1.5$, $\sigma_{AB} = 0.8$, and $\epsilon_{BB} = 0.5$, and $\sigma_{BB} = 0.88$. The dynamics is governed by the Active Ornstein-Uhlenbeck process (AOUP) \cite{Szamel2014-1,Maggi2015-1,fodor2016-1}, which has the following equations:
\begin{equation} \label{particle_pos}
	\boldsymbol{\dot{r}_i}=\xi_0^{-1}\left[\boldsymbol{F_i}+\boldsymbol{f_i}\right],
\end{equation}
\begin{equation}\label{active_force}
	\tau_p \boldsymbol{\dot{f}_i}=-\boldsymbol{f_i}+\boldsymbol{\eta_i}.
\end{equation}
Here $\xi_0$ is the friction that we kept to unity. $\tau_p$ is the persistence time of active particles. $\boldsymbol{f_i}$ is the active force, and $\boldsymbol{F_i}$ is the inter-atomic force, where $i$ is the particle index. The noise-noise correlation is given by $\langle \boldsymbol{\eta_i(t)}\boldsymbol{\eta_j(0)} \rangle = 2 f_o^2 \delta_{ij} \boldsymbol{I}\delta(t)$, where $f_o$ determines the magnitude of the self-propulsion active force acting on the particles. To realize Eqs. (\ref{particle_pos}) and (\ref{active_force}) in simulations, we used the forward Euler scheme in LAMMPS \cite{Thompson2022-1}. We took the time step at least 10 times smaller than $\tau_p$.

To probe the dynamics, we looked into the two-point correlation function $F_s (k, t)$ for the $A$ type particles alone.
\begin{equation}\label{Fskt}
	F_{s}(k,t) = \langle \tilde{f_s}(k,t) \rangle = \frac{1}{N_A}\Big \langle \sum_{i = 1}^{N_A} e^{\dot{\iota} \mathbf{k}.(\mathbf{r}_{i}(t+t_0) - \mathbf{r}_{i}(t_0))} \Big \rangle.
\end{equation}
We chose $k = 7.2$, which is close to the maxima of the static structure factor. The $\langle \ldots \rangle$ includes both time origin averaging and ensemble averaging. The relaxation time $\tau$ is defined as the $1/e$ value of $F_{s}(k,t)$.

The four-point correlation function is defined as the variance of $F_{s}(k,t)$ as
\begin{equation}\label{chi4eq}
	\chi_4(t)=N[\langle \tilde{f_s}(k,t)^2\rangle-F_{s}(k,t)].
\end{equation}

\subsection{Behavior of $\tau$ with $\lambda$}
We show the equilibrium-like behavior of $\tau$ with varing $\lambda$ for different activity. The numerical solution (symbols) of aIMCT agree well with the scaling relation (lines).
\begin{figure}[h]
	\includegraphics[width=13cm]{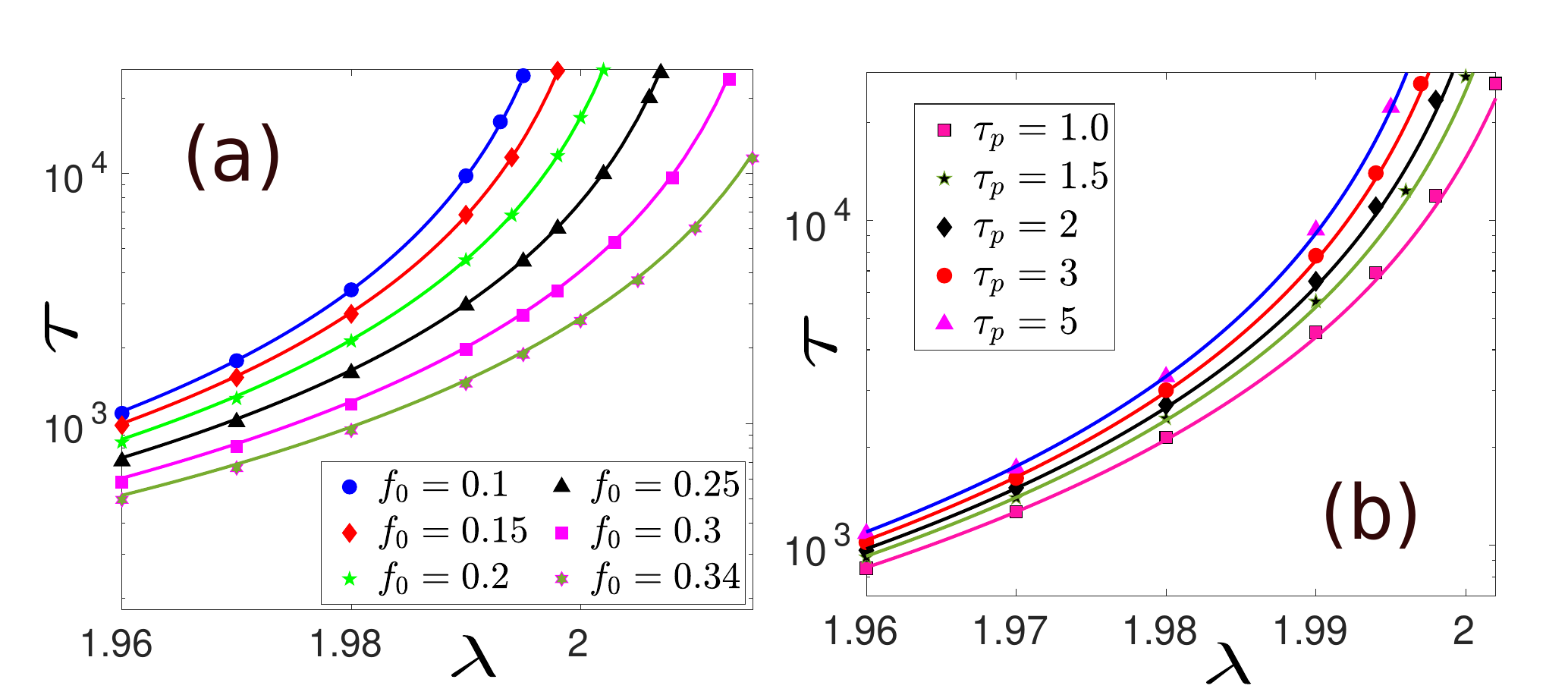}
	\caption{Behavior of $\tau$ with $\lambda$ for different activity. (a) $\tau$ follows the power law prediction of MCT as a function of $\lambda$ for different values of $f_0$. Lines are the fit with effective equilibrium form and points are the MCT results. (b) $\tau$ follows the power law as a function of $\lambda$ also when we vary $\tau_p$. Lines are the fit with effective equilibrium form and points are the MCT results.}
\end{figure}

\subsection{Behavior of $\chi_C^\text{peak}$ with activity for $k_1 = 0.6$ and $k_2 = 2$}
We show the behavior of $\chi_C^\text{peak}$ with varing activity with $k_1 = 0.6$ and $k_2 = 2.0$ which is vaild for schematic solution only. Here we find $\chi_C^\text{peak}$ agrees well with the power-law form considering effective-equilibrium description.

\begin{figure}[h]
	\includegraphics[width=13cm]{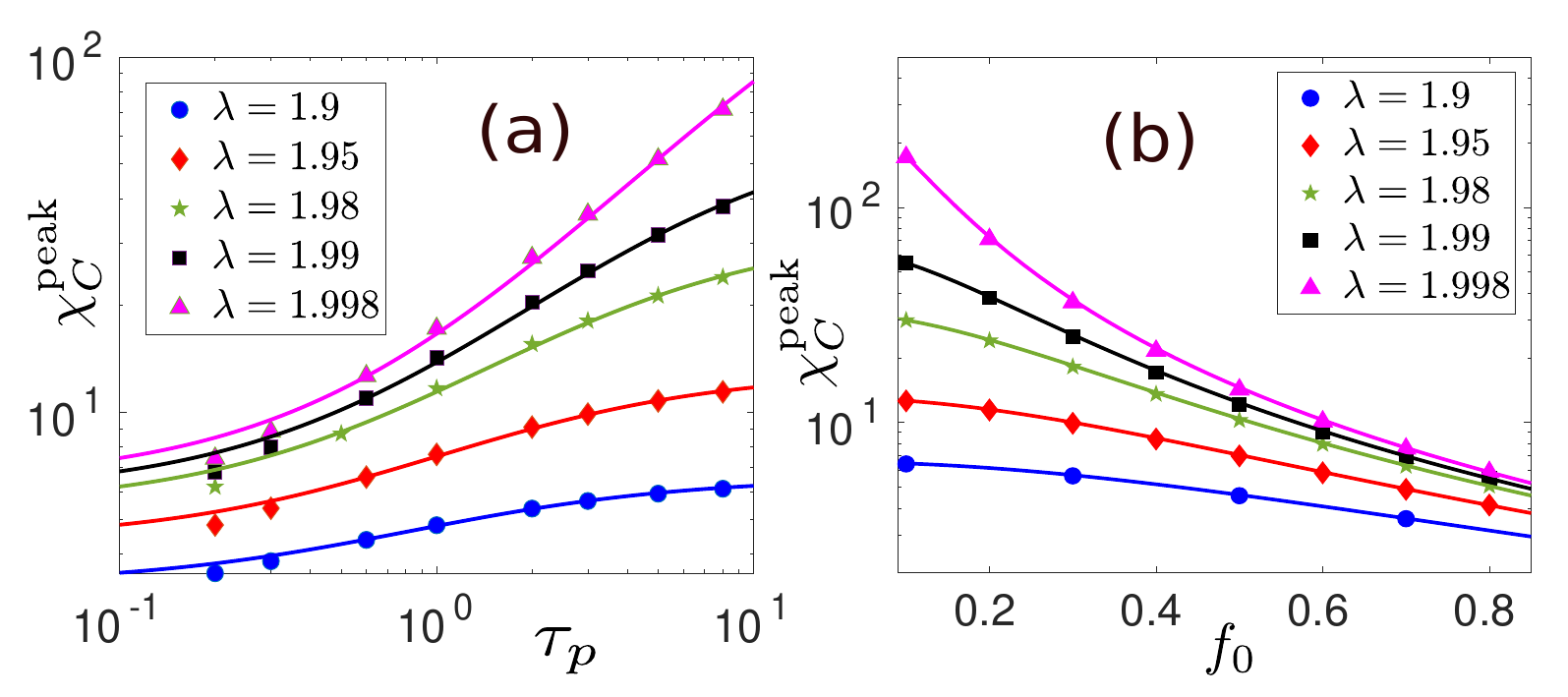}
	\caption{Behavior of $\chi_C^\text{peak}$ with activity. (a) $\chi_C^\text{peak}$ follows the MCT predicted power law with $\tau_p$. Lines are the fit with effective equilibrium form and points are the MCT results. (b) $\chi_C^\text{peak}$ follows the MCT predicted power law with $f_0$. Lines are the fit with effective equilibrium form and points are the MCT results.}
\end{figure}

\end{document}